\documentclass{article}            % STILE

\newcommand{\ket}[1]{|{#1}\rangle} 
\begin{document}

\title{
Basic Logic and Quantum Computing:\\
Logical Judgements by an Insider Observer
}

\author{G.~Battilotti
\\Dipartimento di Matematica Pura ed Applicata, Universita di Padova,\\ 
Via Belzoni 7, 35131 Padova, Italy}

\maketitle
\date{}

\begin{abstract}
We consider the logical assertions of a hypothetical observer who is inside a quantum computer and
 performs a reversible quantum measurement, obtaining a symmetric couple of new axioms, valid only 
inside the quantum computer. The result is that, in this logical framework, symmetry and 
paraconsistency hold.
\end{abstract}

%\keywords{quantum logic; quantum measurement; quantum information} 
\section{Introduction}
Recently, it became clear \cite{DC} that the adequate logical models describing quantum computing 
 are quite different from standard quantum logic \cite{BvN}. In particular, paraconsistency plays 
a relevant role. It seems to us that the deep reason
for this is that standard quantum logic can only describe a quantum measurement at the end of a quantum 
computation, while the computation itself appears as a ''Black Box" to an external observer. The main
aim of our work is to look for the logic of what happens {\em inside} a quantum computer.

In this paper, we make a first step in this direction. We develop the logical interpretation of 
the reversible measurement in quantum computing \cite{Z} in terms of logical assertions, justifying
 the introduction of the logical connectives
interpreting superposition, by exploiting the reflection principle.\cite{SBF} In particular, we show that a particular form of a reversible measurement, called 
''the liar measurement",  can have an interpretation as ''symmetric assertions", that are at the base of a
logical form of symmetry. In this way, logical symmetry from basic logic is justified physically.
The couple of logical axioms we obtain are possible only in a paraconsistent and non-classical 
logical framework, since they are inconsistent with the non-contradiction and with the excluded middle
principles, as it is also suggested by the models in Refs. \cite{DC}, \cite{DCG}.

\section{Reversible Measurement and Logical Assertions}

It is well known that the information unit in quantum computing is given by the {\em qubit}, 
mathematically represented by the vector 
\begin{equation}\label{qubit}
     q=a\ket{0}+b\ket{1},
\end{equation}
where $a, b$ are complex numbers 
(called probability amplitudes) and $|a|^2 + |b|^2 = 1$.
Such information can be partially extracted by a standard quantum measurement, which can find 
$\ket{0}$
(with probability $|a|^2$) or $\ket{1}$ (with probability $|b|^2$). Then the whole information 
contained in $q$ is hidden, since after the measurement the qubit collapses into one of the two states
$\ket{0}$ or $\ket{1}$, and superposition is lost. 

As it is well known (cf. \cite{NC}, pg. 187), the
measurement can be performed in any orthonormal basis of $\mathbf{C}^2$. Let $A$, $A^\perp$ be a couple of 
orthonormal states. After a measurement in such a basis, an observer can assert 
\centerline{''the qubit is in the state $A$",}
or 
\centerline{''the qubit is in the state $A^\perp$".} 
\smallbreak
\noindent We write such assertions by the notations 
$$
\vdash A
$$ and
$$
\vdash A^\perp
$$ respectively. Here we borrow Frege's notation of truth, meaning in our case 
that ''$A$ ($A^\perp$) is true because it is the result of a measurement". We have used the same 
notation in Ref. \cite{SBF} consistently with the notion of sequent, but, in this  setting,  the 
meaning of sequent, that is of logical consequence,
is still an open problem and will be object of future work.

Obviously, the observer can assert {\em only one} of the two assertions, since, after the standard 
measurement, the superposition is lost and cannot be recovered any more (for the geometrical meaning of 
this fact see Ref. \cite {Z}). Then a further measurement applied after the previous one would be 
irrelevant,
being a classical measurement of one bit.

A different case  would arise if the observer could apply two measurements in parallel. Notice,
 however, 
that we cannot perform two standard quantum measurements in parallel on two identical copies of our
qubit $q$, because of the no-cloning theorem.\cite{WZ} The only way to achieve this is to perform a 
new kind of quantum measurement (called {\em basic} or {\em reversible} measurement) on the same qubit 
$q$, as showed in Ref. \cite {Z}. We remind that the basic measurement consists of the observation of the
qubit made by an hypothetical internal observer, who lives inside the quantum computer, or, which is 
the same, in a quantum space whose states are in a one-to-one correspondence with the computational
states. The action of the insider observer is in fact the application of a particular quantum gate, 
which is the linear superposition of two orthogonal projectors, each one corresponding to a standard 
quantum measurement. As quantum gates are unitary operators,
the basic measurement is reversible, then the quantum information is not hidden any more.    
In this way, by the reversible measurement, the internal observer 
can obtain the two judgements together:
\begin{equation}\label{coppia1}
\vdash A \qquad\qquad\qquad \vdash A^\perp
\end{equation}

We remind that, following the reflection principle \cite{SBF}, the logical connectives are the 
result of 
importing into the formal language some pre-existent metalinguistic links between assertions.  
Let us consider
the physical link of superposition between orthogonal states, which is possible only inside the
quantum computer. This becomes a logical link between opposite judgements once the superposition has been measured,
obtaining the couple of judgements (\ref{coppia1}). By the reflection principle, we interpret the 
justaposition
of two judgements as the additive conjunction $\&$, putting: $\vdash A\& B \equiv \; \vdash 
A \quad \vdash B$. 
In our case, we put:
\begin{equation}
\vdash A\& A^\perp \equiv \;\vdash A \quad \vdash A^\perp
\end{equation}
So the reversible measurement gives back the superposed judgement
\begin{equation}\label{ax1}
\vdash A\& A^\perp
\end{equation}
that is, it makes true the superposition of the two orthogonal states. The proposition 
''$A \& A^\perp"$,
which is judged true, is a qualitative logical way to represent superposed states,
without mentioning the probability amplitudes.

This result should be considered as an axiom for any logical system  adequate to quantum computation.
We stress that this axiom is derived from a thought experiment (the reversible measurement) and not
only by logical reasoning. This axiom states that, in the case of judgements derived from two  orthogonal 
states of a quantum computer, ''contradiction" is true, so that the non-contradiction principle cannot 
hold inside the quantum computer. This suggests that the adequate logic for a quantum computer should
be paraconsistent. 

\section{The Liar Measurement and the Falsity Judgements}

Let us suppose now that the external observer applies a standard quantum measurement of $q$ in the 
orthonormal basis $A,A^\perp$ and then she decides 
to apply a NOT gate of a classical computer to the result of her measurement. 
If she had obtained for example $A$ after the measurement, she would get $A^\perp$ after the NOT, but
now she cannot assert her final result as true! In fact, the correct assertion is that $A^\perp$ is false. 
We write such assertion as in Ref. \cite{SBF}:
$$
A^\perp \vdash
$$
that is a {\em primitive} way to say that $A^\perp$ is false.
Then the ''NOT-measurement" just discussed, reversed the original judgement: $(\vdash A)^\perp$.
So we put the equivalence:
$$
(\vdash A)^\perp \equiv A^\perp\vdash.
$$
If instead she measured $A^\perp$ and performed the negation afterwords, she would get the 
judgement:
$$
A\vdash
$$
obtained by reversing the original judgement $(\vdash A^\perp)^\perp$. 
So we put the second equivalence:
$$
(\vdash A^\perp)^\perp \equiv A\vdash
$$
If she were able to make the two ''NOT-measurements" together, she would obtain both:
$$ 
A^\perp\vdash\qquad\qquad A\vdash
$$
This is  a logical link between two ''falsity judgements" which, as in Ref. \cite{SBF}, is solved as follows:
$$
A^\perp \oplus A\vdash\; \equiv \;A^\perp\vdash\quad A\vdash
$$
where $\oplus$ is the additive logical disjuncion.
Of course, the external observer cannot do so, always for the same reasons. But the internal observer can 
perform the two NOT-measurements at the same time, as she also has the NOT gate! In fact, she gets 
the so called ''liar
measurement'' which is just the NOT gate applied after a basic measurement (for more detail, see Ref. 
\cite{BZ}). After the liar measurement, the internal observer can still see a superposed state,
but with the original amplitudes $a$ and $b$ interchanged. For this reason the prefix ''Liar''
was added, in order to stress the fact that the external observer is cheated when she performs
a standard quantum measurement after a liar measurement.

In logical terms, the liar measurements means that inside the quantum computer the judgement 
$\vdash A\& A^\perp$ has been negated:
$$
(\vdash A\& A^\perp)^\perp\equiv A^\perp \oplus A\vdash
$$
The falsity judgement 
\begin{equation}\label{ax2}
A^\perp \oplus A\vdash 
\end{equation}
is the second axiom obtained from a reversible measurement.
It states that the excluded middle principle does not hold inside the quantum computer.  The interesting
fact is that the second axiom (\ref{ax2}) could be recovered simply by symmetry, as in basic 
logic,\cite{SBF} from the first axiom (\ref{ax1}). 
In Ref. \cite{BZ} we will show that this logical symmetry has a truly geometrical
origin.

\section*{Aknowledgments}
I am very grateful to Paola Zizzi for many useful discussions.

Work supported by the research project ''Logical Tools for Quantum Information Theory'', 
Department of Mathematics, University of Padova, Italy.

%%%%%%%%%%%%%%%%%%%%%%%%%%%%%%%%%%%%%%%%%%%%%%%%%%%%%%%%%%%%%%%%%%%%%%%%%%%

\end{document}